\documentclass[english,letterpaper,twocolumn,showpacs,pra,aps]{revtex4}
\usepackage{graphicx}
\usepackage{amssymb}

\makeatletter

\large  

\input epsf

\makeatother

\usepackage{babel}
\makeatother
\begin{document}

\title{Interplay of Aharonov-Bohm, chirality, and aspect ratio effects in the axial conductance of a nanotube}

\author{Eugene B. Kolomeisky$^{1}$, Hussain Zaidi$^{1}$ and Joseph P. Straley$^{2}$}

\affiliation
{$^{1}$Department of Physics, University of Virginia, P. O. Box 400714,
Charlottesville, Virginia 22904-4714, USA\\
$^{2}$Department of Physics and Astronomy, University of Kentucky,
Lexington, Kentucky 40506-0055, USA}

\begin{abstract}
A magnetic field applied along the axis of a nanotube can counteract the effect of the tube chirality and dramatically affect its conductance, leading to a way to determine the chirality of a nanotube. The effect of the applied field is strongest in the long tube limit where the conductance is (i) either a sequence of sharp $4e^{2}/h$ height peaks located at integer values of the flux (for an armchair tube) or (ii) a periodic sequence of pairs of $2e^{2}/h$ height peaks for a chiral tube, with the spacing determined by the chirality.  In the short tube limit the conductance takes on the value that gives the universal conductivity of an undoped graphene sheet, with a small amplitude modulation periodic in the flux.  
\end{abstract}


\maketitle
\section{Conductance of graphene sheet and nanotubes}
Katsnelson \cite{Katsnelson} has given a very general and elegant explanation of the experimental observation that at zero temperature an ideal graphene sheet has conductivity of the order $e^{2}/h$.  Although earlier considerations \cite{RMP} arrived at essentially the same conclusion, Katsnelson's argument stands out because it clearly demonstrates that the finite conductivity of graphene is a direct consequence of the electron dynamics, which is described by the two-dimensional massless Dirac equation.  Katsnelson \cite{Katsnelson} has considered an undoped graphene sheet in the shape of a cylinder of length $L$ and circumference $W$ (with both length scales significantly larger than the lattice spacing of graphene) attached at its bases to heavily doped graphene leads and demonstrated by means of the Landauer transmission formula \cite{Landauer} that the axial conductivity of the cylinder in the ring limit, $W/L\rightarrow\infty$, approaches the universal $4e^{2}/\pi h$ value.  The crucial element of the analysis is an expression for the transmission probability of propagating modes in the leads \cite{Katsnelson}
\begin{equation}
\label{transmission}
T_{n}=\frac{1}{\cosh^{2}(q_{n}L)}
\end{equation}  
where the index $n$ labels the modes, and the azimuthal wave numbers $q_{n}=2\pi n/W, n=0, \pm 1, \pm 2, ...$ are singled out by the condition of periodicity around the circumference.  Then, accounting for the spin degeneracy, the conductance of the cylinder (hereafter measured in units of the conductance quantum $2e^{2}/h$) per Dirac valley $K$ (or $K'$) \cite{RMP} is given by
\begin{equation}
\label{no_flux_conductance}
G_{K}=\sum_{n=-\infty}^{\infty}T_{n}=\sum_{n=-\infty}^{\infty}\frac{1}{\cosh^{2}(2\pi nL/W)}
\end{equation}
Accounting for the presence of the two degenerate Dirac valleys, the total conductance is $G=2G_{K}$ and the conductivity $\sigma$ follows from the relationship $\sigma = GL/W$. 

Tworzyd{\l}o \textit{et al.} \cite{Tworzydlo} extended Katsnelson's analysis to the case of an open strip geometry (length $L$ and width $W$) and calculated both the conductivity and the Fano factor of the shot noise.  They found the same expression (\ref{transmission}) for the transmission probability except that the spectrum $q_{n}$ was now determined by the boundary conditions at the edges of the strip.  Assuming that longitudinal and transverse momenta are not coupled, a general conductance formula that summarizes various boundary conditions and accounts for all degeneracies can be given in the form \cite{Tworzydlo}
\begin{equation}
\label{strip_conductance}
G_{strip}=\sum_{n=-\infty}^{\infty}\frac{1}{\cosh^{2}[(n+\alpha)\pi L/W]}
\end{equation}
with $\alpha=1/2$ for a an edge defined by a slowly varying potential \cite{Berry}, and $\alpha=0$ or $\alpha=1/3$ for an abrupt armchair edge, depending on geometric details of the strip \cite{Tworzydlo,Fertig}.   In the short and wide strip limit, $W/L\rightarrow \infty$, Tworzyd{\l}o \textit{et al.}, arrived at the same universal conductivity as Katsnelson \cite{Katsnelson} but also observed that the boundary conditions have a significant effect in the long and narrow strip limit ($W/L\rightarrow 0$):  the smooth boundary ($\alpha=1/2$) yields an insulator, while a boundary giving $\alpha=0$ is a conductor.    We will discuss below what happens for general $\alpha$ and aspect ratio.  

In addition to this sensitivity to boundary conditions in the strip geometry, there is an analogous sensitivity to the chirality of the graphene cylinder, i. e. to the way the graphene sheet is wrapped into a cylinder.  Indeed, as written,  Eq.(\ref{no_flux_conductance}) predicts that an infinitely long tube ($W/L=0$) possesses unit conductance per valley, i. e. it is a semimetal.  However it is known that only about one-third of all possible tubes are semimetallic.  It turns out that Eq.(\ref{no_flux_conductance}) can be generalized to account for all kinds of tubes without affecting the main conclusion of Ref.\cite{Katsnelson}.  The generalization is due to Ajiki and Ando \cite{Ando} and Kane and Mele \cite{Kane} who observed that the long-wavelength low-energy dynamics of the graphene electrons of the $K$-valley when confined to cylindrical surface is described by the two-dimensional massless Dirac equation in the presence of a fictitious vector potential that represents the effects of the tube size and its chirality.  The effect is to replace the azimuthal quantum number $n$ in Eq.(\ref{no_flux_conductance}) with $n+\alpha$ thus making the outcome
\begin{equation}
\label{no_flux_chiral_conductance}
G_{K}(\alpha)=\sum_{n=-\infty}^{\infty}\frac{1}{\cosh^{2}[(n+\alpha)2\pi L/W]}
\end{equation}
similar to Eq.(\ref{strip_conductance}), but now $G_K$ vanishes for large $L/W$. An expression very similar to Eq.(\ref{no_flux_chiral_conductance}) is also encountered in the analysis of the conductance of a planar graphene flake having Corbino disc topology \cite{Corbino}. 

The intrinsic flux $\alpha$ combines the effects of winding and curvature, and provides a classification of nanotubes \cite{Ando,Kane}.  In terms of the usual $(M,N)$ classification of nanotubes (which explains how the graphene structure is wrapped around the cylinder), $\alpha$ has a geometrical interpretation.   The new periodicity introduced into the graphene structure in forming the nanotube limits the accessible reciprocal lattice vectors to lines which in general do not visit the Dirac points where the gap vanishes; $\alpha$ measures the distance between Dirac point and line in units of the periodicity wavevector $k_{\perp} = 2 \pi/W$.  When $N-M$ is not a multiple of 3, $\alpha$ is close to $\frac {1}{3}$ (insulating nanotube); otherwise
 \cite{Kleiner} the nanotubes are semiconductors with
\begin{equation}
\label{alphaexpression} 
\alpha = \frac {\sqrt{3} \pi}{48} \frac {(N-M)(2 N^2 + 5 N M + 2 M^2)}{(N^2 + NM + M^2)^2}
\end{equation}
The largest values occur for $M=0$ ("zigzag" nanotubes); for $M=N$, $\alpha = 0$ ("armchair" nanotubes, which are semimetals).
 
The fictitious flux has opposite signs in the two Dirac valleys \cite{Ando} because the flux is not due to a physical magnetic field and the system still has time-reversal symmetry. The contribution of the electrons of the $K'$-valley into the conductance, $G_{K'}(\alpha) = G_{K}(-\alpha)$, is still given by Eq.(\ref{no_flux_chiral_conductance}) because it is an even function of chirality $\alpha$.  Thus the net conductance is given by $G=2G_{K}(\alpha)$.      

\section{Effects of an axial magnetic field}

Since the chirality enters the problem as a fictitious flux, the physical properties of graphene tubes can be dramatically affected by the presence of an axial magnetic field as was anticipated theoretically \cite{Ando,Datta,Roche} and observed experimentally \cite{experiment}.    Indeed, imagine there is an axial Aharonov-Bohm (AB) flux $\Phi$ \cite{AB} entirely confined within the interior of the cylinder.   Adopting the gauge where the vector potential points in the azimuthal direction, we observe that the electrons experience a vector potential $A=\Phi/W$ whose effect can be accounted for via the Peierls substitution $q_{n}\rightarrow q_{n} +eA/\hbar c= (2\pi/W)(n+\phi)$ where $\phi=\Phi/\Phi_{0}$ is the dimensionless  flux measured in units of the flux quantum $\Phi_{0}=hc/e$.  In present geometry the Peierls substitution is exact because the vector potential has constant magnitude and direction in the natural coordinate system of the tube.  Then, in the light of the above discussion, the net conductance of the cylinder of chirality $\alpha$ in the presence of the AB flux $\phi$ will be given by the sum of contributions due to two Dirac valleys
\begin{equation}
\label{net_conductance_chirality_flux}
G(\phi, \alpha)= G_{K}(\phi+\alpha)+G_{K}(\phi-\alpha)
\end{equation}
We note that in contrast to the fictitious flux $\alpha$, the AB flux $\phi$ breaks the time-reversal symmetry thus lifting the degeneracy between the $K$- and $K'$-valleys \cite{Ando}.  The conductance (\ref{net_conductance_chirality_flux}) is an even function of both $\phi$ and $\alpha$.   The interesting implication is that an applied magnetic field can cancel the effect of chirality, restoring conductivity to a nonmetallic nanotube.

The conductance of an infinitely long tube ($W/L=0$) has been discussed by Tian and Datta \cite{Datta}.   In this case the effect of the flux and chirality on the conductance of a tube can be anticipated by looking at the square of the energy eigenvalue of the Dirac equation $E_{n}^{2}(q_{z})=\hbar^{2} v^{2}[q_{z}^{2}+(2\pi/W)^{2}(n+\phi\pm\alpha)^{2}]$ where $v$ is the Fermi velocity, $q_{z}$ is the wave vector in the axial direction, and the upper and lower signs refer to the $K$- and  $K'$-electrons.  Since the integer part of $\alpha$ or $\phi$ can be absorbed into the definition of the azimuthal quantum number $n$, any measurable property will be a periodic function of $\alpha$ or $\phi$ with unit period.  Then the energy gaps 
\begin{equation}
\label{gaps}
\Delta_{K,K'}=\frac{4\pi \hbar v}{W}|\phi\pm\alpha|
\end{equation}
can be periodically closed by the AB flux $\phi$ thus triggering semiconductor-semimetal transitions \cite{Ando,Datta}.  The way these transitions manifest themselves in the flux dependence of the conductance is sensitive to the value of $\alpha$.  The gaps in the two Dirac valleys (\ref{gaps}) can be  closed simultaneously only if $\alpha \equiv 0$ (the case $\alpha=1/2$ does not occur in nanotubes).  If  both valleys become ungapped at once ($\Delta_{K,K'}=0$), there are two open ideal conductive channels, and the conductance  (according to the Landauer formulation \cite{Landauer,Datta}) is $4e^{2}/h$, i.e. twice the conductance quantum \cite{ballistic_1}.  These conductance peaks are located at integer \cite{Datta} ($\alpha=0$) values of the AB flux $\phi$.  In all other cases, the AB flux $\phi$ can only close one of the gaps (\ref{gaps}) at a time.  As a result there is only one conductive channel present and the conductance is given by the conductance quantum $2e^{2}/h$ \cite{ballistic_2}.  These peaks appear in pairs and the pair with its center at integer flux has a separation of $\Delta\phi=2|\alpha |$.  For the $2e^{2}/h$ conductance peak, the axial current is entirely due to the electrons of the single open valley.

This leads to the possibility of "$\alpha$ spectroscopy," where the magnetic field response can help index a nanotube.  Unfortunately, this is somewhat constrained by available magnetic fields. The flux quantum is $4.138 \times 10^{5}$ $Tesla\cdot\AA^{2}$, with the result that the size of the magnetic field $B$ needed to see one complete AB cycle for an extremely large tube of radius $R=25 \AA$ is 200 Tesla \cite{Ando}.  This is still a very small field compared to characteristic field $B^{*}$ of onset of disruption of normal atomic structure.   The field $B^{*}$ can be estimated from the condition of equality of the magnetic length $\sqrt{(\hbar c/eB)}$ to the Bohr radius $\hbar{^2}/me^2$ or equivalently requiring that there is one flux quantum per cross-sectional area of undisturbed atom. This gives the field $B^{*} \simeq m^{2}e^{3}c/\hbar^{3}$ which is over $10^{5}$ Tesla.

When $N-M$ is not a multiple of 3,  nanotubes are insulating and have $\alpha$ close to $1/3$.  Achieving $\phi = 1/3$ (i.e. having $B \times Area = \frac {1}{3} \Phi_{0}$) is also not possible with the available magnetic fields and nanotubes of experimental interest.    However, according to (\ref{alphaexpression}), $\alpha$ can be significantly smaller when $N-M = 0$ (mod 3).   The $M=N$ "armchair" nanotubes have $\alpha = 0$, and for other cases the curvature of the tube surface gives $\alpha \simeq 0.23/N\simeq 0.09 \AA/R$ for "zigzag" tubes ($M = 0$), and $\alpha \simeq 1.9/N^{2} \simeq 0.16 \AA^{2}/R^{2}$ for the "nearly armchair" case $M=N-3$.  For this latter case, the magnetic field needed to achieve conductance at low temperatures is $B = 10^{4} T \cdot\AA^4/R^{4}$, implying that a 11 Tesla field would be adequate for the $(8,11)$ nanotube ($R=6.5 \AA$; the gap would be $0.006 eV$). Although the "armchair" nanotubes are conducting at zero magnetic field, application of a longitudinal magnetic field would render them nonconducting. 

\section{Effects of finite length}

The sharpness of the semiconductor-semimetal transitions discussed above is determined by the aspect ratio $W/L$.  Consider the case where there is no applied magnetic field.  In the ring limit ($W/L\rightarrow\infty$), the sum in Eq.(\ref{no_flux_chiral_conductance}) can be approximated by an integral with the result $G_{K}(\alpha)\simeq(1/\pi)(W/L)$ which leads to a finite universal conductivity $\sigma_{0}=2G_{K}(\alpha)L/W=2/\pi$ \cite{Katsnelson,Tworzydlo}.  This is chirality-independent because the quantization of the azimuthal motion is irrelevant in the ring limit.   

In the tube limit ($W/L\rightarrow0$),  the discreteness of the transverse wavevectors is relevant.  The sum in Eq.(\ref{no_flux_chiral_conductance}) is well approximated by the $n=0$ and $n=-1$ terms (assuming $0 < \alpha \leqslant 1/2$):  
\begin{equation}
\label{tube_conductance}
G(0,\alpha)\simeq\frac{2}{\cosh^{2}(2\pi \alpha L/W)}+ \frac{2}{\cosh^{2}[(1-\alpha)2\pi L/W]}
\end{equation}
In the  case $\alpha=0$, this simplifies to $G(0,0)\simeq2+8\exp(-4\pi L/W)$
where the first term represents the conductance of an infinite tube and the second term is the correction due to finite aspect ratio.  The conductivity $\sigma=GL/W\simeq L/W$ is infinite.  Since in the ring limit ($W/L\rightarrow\infty$) the conductivity approaches $2/\pi$, the conductivity is a monotonically decreasing function of the aspect ratio for $\alpha = 0$, as shown by Tworzyd{\l}o \textit{et al.} \cite{Tworzydlo} in their analysis of Eq.(\ref{strip_conductance}) for $\alpha=0$.

Turning on finite chirality or magnetic field, no matter how small, has a dramatic effect on the conductance of a long tube.  For $0<\alpha<1/2$ the first term of Eq.(\ref{tube_conductance}) dominates and we find exponentially small conductance $G(0,\alpha)\simeq4\exp(-4\pi\alpha L/W)$; the tube is an insulator with a gap of order $\alpha R$.  We observe that the conductivity $\sigma(0,\alpha)\simeq4(L/W) \exp(-4\pi \alpha L/W)$ has a maximum for $W/L\simeq4\pi\alpha$.  As $\alpha\rightarrow0$, the conductivity maximum shifts towards $W/L=0$ and becomes larger.  For $\alpha=1/2$ both terms of (\ref{tube_conductance}) contribute equally and we find again both vanishing conductance and vanishing conductivity.  As shown in Figure 1, there is a characteristic chirality  $\alpha_{c} = 1/4$ beyond which there is no maximum at all and the conductivity is a monotonically increasing function of the aspect ratio $W/L$.  
\begin{figure}
\includegraphics[width=1.0\columnwidth, keepaspectratio]{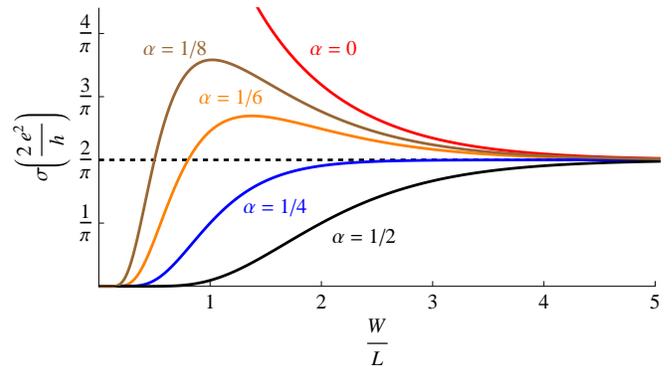} 
\caption{(Color online) Conductivity versus aspect ratio $W/L$ for different tube chiralities parameterized by the fictitious flux $\alpha$ as given by Eq.(\ref{Poisson_chirality_conductivity}).}
\end{figure} 

According to Figure 1, the tube limit corresponds to $W/L < 2 \alpha$.  This will give a nonconducting tube at low temperatures.  This also describes the sensitivity to applied magnetic field:  when $|\phi-\alpha| > 2W/L$ the conductance will be suppressed, while for $|\phi - \alpha| <  W/\pi L$ the conductance will be close to the ideal value $G \simeq 1$.  For a $1$ micron-long nanotube with circumference $30 \AA$ this means that the flux has to be set to within $0.3 \%$ to observe restoration of conductance (if $\alpha \ne 0$); but this means it will take a $16 T$ field to render an "armchair" nanotube of these dimensions nonconducting.  We note that currently available nanotubes have aspect ratios in the $W/L< 0.79$ range \cite{short}.

The same Figure gives the conductivity of the graphene strip with the aspect ratio of $2L/W$ (compare Eqs. (\ref{strip_conductance})-(\ref{net_conductance_chirality_flux})). The $\alpha=0$ and $\alpha=1/2$ plots were previously given \cite {Tworzydlo}.

When $W/L$ is large it is useful to use the Poisson summation formula \cite{LL5} to the sum over $n$ in Eq.(\ref{no_flux_chiral_conductance}) with the results for the net conductance
\begin{equation}
\label{Poisson_chirality_conductance}
G(0,\alpha)=2G_{K}(\alpha)=\frac{W^{2}}{L^{2}}\sum_{n=-\infty}^{\infty}\frac{n\cos(2\pi n\alpha)}{\sinh(\pi nW/2L)}
\end{equation}
and conductivity
\begin{equation}
\label{Poisson_chirality_conductivity}
\sigma(\alpha)=\frac{2}{\pi}+ \frac{2W}{L}\sum_{n=1}^{\infty}\frac{n\cos(2\pi n\alpha)}{\sinh(\pi nW/2L)}
\end{equation}
respectively.  We now observe that for \textit{arbitrary} aspect ratio $W/L$ the effect of chirality manifests itself in periodic oscillations of the conductivity about $\sigma_{0}=2/\pi$;  the aspect ratio determines the amplitude of the conductivity oscillations.  

In the ring limit, $W/L\rightarrow\infty$, the sum in Eq.(\ref{Poisson_chirality_conductivity}) is well-approximated by the constant and $n=1$ terms with the result
\begin{equation}
\label{Poisson_flux_conductivity_ring}
\sigma(\alpha)\simeq\frac{2}{\pi}+ \frac{4W}{L}\exp \left (-\frac{\pi W}{2L}\right )\cos(2\pi\alpha)
\end{equation}
We observe that as $W/L\rightarrow\infty$ the limiting conductivity of $2/\pi$ is approached from above if $0\leqslant\alpha<1/4$ and from below if $1/4\leqslant\alpha\leqslant1/2$.  This demonstrates that $\alpha_{c}=1/4$ as suggested by Figure 1.  

For the case when there is an applied field the Poisson transformation can again be used to give a representation more suitable to finite $W/L$:  
\begin{equation}
\label{Poisson_conductance_chirality_flux}
G(\phi,\alpha)=\frac{W}{L} \left (\frac{2}{\pi}+ \frac{2W}{L}\sum_{n=1}^{\infty}\frac{n\cos(2\pi n\phi)\cos(2\pi n\alpha)}{\sinh(\pi nW/2L)} \right )
\end{equation}

We now see that the combined effects of the AB flux $\phi$ and chirality $\alpha$ manifest themselves in oscillations (periodic in both $\phi$ and $\alpha$)  of the conductance about the value of $2W/L\pi$ corresponding to the universal conductivity of graphene $\sigma_{0}=2/\pi$;  the shape and amplitude of the oscillation is determined by the aspect ratio $W/L$.  In the ring limit $W/L\rightarrow \infty$ the conductance is dominated by the constant term of Eq.(\ref{Poisson_conductance_chirality_flux}) and the leading finite aspect ratio correction (if $\alpha \neq 1/4$) is supplied by the $n=1$ term:
\begin{equation}
\label{Poisson_conductance_flux_chirality_ring}
G(\phi,\alpha)\simeq \frac{2W}{\pi L}+ \frac{4W^{2}}{L^{2}}\exp \left (-\frac{\pi W}{2L}\right )\cos(2\pi \phi)\cos(2\pi \alpha)
\end{equation}
Our results summarizing the effects of chirality and AB flux on the conductance of semimetallic ($\alpha=0$), insulating ($\alpha=1/3$) and semiconducting (mimicked by $\alpha=1/9$) tubes are shown in Figure 2.
\begin{figure}
\includegraphics[width=1.0\columnwidth, keepaspectratio]{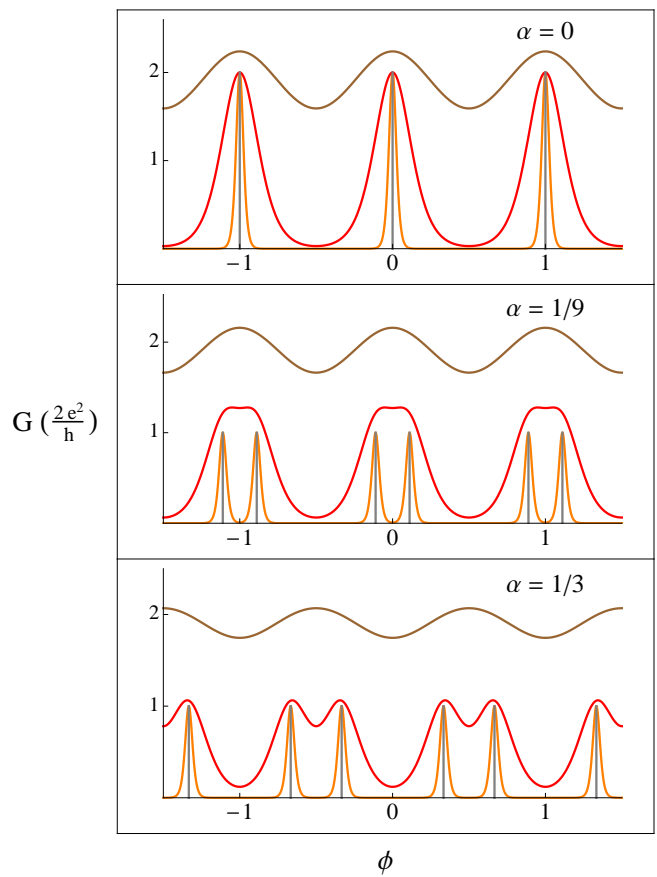} 
\caption{(Color online) Conductance versus dimensionless AB flux $\phi$ for different tube chiralities parameterized by fictitious flux $\alpha$ and aspect ratio $W/L$ as given by Eq.(\ref{Poisson_conductance_chirality_flux}).  Aspect ratios are color-coded according to $W/L=0$ (grey), $W/L=1/5$ (orange), $W/L=1$ (red), and $W/L=3$ (brown).}
\end{figure} 

Since the conductance represents directly measurable quantity, we hope that our conclusions will be experimentally tested.  The chirality of a tube can be determined by measuring the dependence of the conductance on the AB flux of a not very short cylinder, and a conducting "armchair" tube can be rendered nonconducting.  

The issue of separation of the effects we predicted from those due to the leads is an important one.  First we note that there is a series of conductance measurements \cite{experiment,ballistic_1,ballistic_2,short}  and an expertise on how to deal with the end effects is growing.  For example, Javey \textit{et al.} \cite{short} found that the resistance of a nanotube was rather close to the expected $6.5$ $k\Omega$; they estimated that the leads and contacts contributed no more than $0.6$ $k\Omega$ (and much less at low temperatures).  We predict that the tube will go from conducting to nonconducting (or the other way around) in a magnetic field of specified size (the field does not have to be aligned but the part that is not axial will not contribute).   The end effects are in series with this; it will complicate telling what the conductance of the undisturbed part of the tube is, but the change in conductance should be straightforward to observe.  

The conductance could also be monitored without contacts.   An electric field $E$ parallel to the axis of a conducting nanotube of length $L$ will give a dipole moment of order  $L^{3} E$ (in Gaussian units).   Then light polarized parallel to the axis of a bundle of oriented nanotubes will be scattered, provided that its frequency $\omega$ is low enough for the charge separation to take place.   For a nanotube with the Landauer conductance $2 e^{2}/h$, this means $\omega < 10^{6}Hz\cdot m^{2}/L^{2}$  or $\omega < 10^{18} Hz$ for micron-length tubes.   But if the conductance is suppressed by a magnetic field, so is the rate of charge transport, and then there is much less scattering at intermediate frequencies.

\section{Acknowledgments}

We thanks D. R. Strachan for sharing with us his experimental expertise.  This work was supported by the US AFOSR grant FA9550-11-1-0297.

\end{document}